\documentclass[prl,aps,amsmath,amssymb,amsfonts,twocolumn,nofootinbib,showpacs]{revtex4}
\usepackage{bm,mathrsfs}

\usepackage{graphicx}
\usepackage{epsfig}
\usepackage{amsmath,bbm}
\usepackage{amsfonts,amssymb}
\usepackage{times}
\usepackage{verbatim}
\usepackage[sort&compress]{natbib}

\usepackage{amsmath}
\usepackage[colorlinks,breaklinks,linkcolor=blue,anchorcolor=blue,citecolor=blue,urlcolor=blue,
dvipdfm
]{hyperref}

\newcommand{\be}{\begin{equation}}
\newcommand{\ee}{\end{equation}}
\newcommand{\bea}{\begin{eqnarray}}
\newcommand{\eea}{\end{eqnarray}}

\newcommand{\ket}[1]{|#1\rangle}
\newcommand{\bra}[1]{\langle#1|}

\def\>{\rangle}
\def\<{\langle}

\def\qed{\leavevmode\unskip\penalty9999 \hbox{}\nobreak\hfill
     \quad\hbox{\leavevmode  \hbox to.77778em{%
               \hfil\vrule   \vbox to.675em%
               {\hrule width.6em\vfil\hrule}\vrule\hfil}}
     \par\vskip3pt}
     
\begin{document}

\newtheorem{theorem}{Theorem}
\newtheorem{lemma}[theorem]{Lemma}
\newtheorem{corollary}[theorem]{Corollary}
\newtheorem{proposition}[theorem]{Proposition}
\newtheorem{definition}[theorem]{Definition}
\newtheorem{example}[theorem]{Example}
\newtheorem{conjecture}[theorem]{Conjecture}

\title{{Dissipation-based entanglement via quantum Zeno dynamics and Rydberg antiblockade }}

\author{X. Q. Shao}
\affiliation{Center for Quantum Sciences and School of Physics, Northeast Normal University, Changchun, 130024, People's Republic of China}
\affiliation{Center for Advanced Optoelectronic Functional Materials Research, and Key Laboratory for UV Light-Emitting Materials and Technology
of Ministry of Education, Northeast Normal University, Changchun 130024, China}

\author{J. H. Wu}
\affiliation{Center for Quantum Sciences and School of Physics, Northeast Normal University, Changchun, 130024, People's Republic of China}
\affiliation{Center for Advanced Optoelectronic Functional Materials Research, and Key Laboratory for UV Light-Emitting Materials and Technology
of Ministry of Education, Northeast Normal University, Changchun 130024, China}

\author{X. X. Yi}
\affiliation{Center for Quantum Sciences and School of Physics, Northeast Normal University, Changchun, 130024, People's Republic of China}
\affiliation{Center for Advanced Optoelectronic Functional Materials Research, and Key Laboratory for UV Light-Emitting Materials and Technology
of Ministry of Education, Northeast Normal University, Changchun 130024, China}
\date{\today}

\begin{abstract}
A novel scheme is proposed for dissipative generation of maximally entanglement between two Rydberg atoms
in the context of cavity QED. The spontaneous emission of atoms combined with quantum Zeno dynamics and Rydberg antiblockade guarantees a unique steady solution of the master equation of system, which just corresponds to the antisymmetric Bell state $|S\rangle$. The convergence rate is accelerated by the ground-state blockade mechanism of Rydberg atoms. Meanwhile the effect of cavity decay is suppressed by the Zeno requirement, leading to a steady-state fidelity about $90\%$ as the single-atom
cooperativity parameter $C\equiv g^2/(\kappa\gamma)= 10$, and this restriction is further relaxed to  $C= 5.2$ once the quantum-jump-based feedback control is exploited.
\end{abstract}

\pacs{03.67.Bg,32.80.Ee,42.50.Dv,42.50.Pq}

\maketitle


The quantum dissipation arising from weak coupling between a quantum system and its surrounding reservoirs, is usually indicated by a Lindblad generator in Markovian quantum master equations for open quantum systems. Compared with unitary dynamics, the dissipative dynamics does provide a more accurate image for
characterizing the evolution of quantum states in a realistic situation. It has always been regarded that the quantum dissipation plays a negative role in quantum information processing (QIP) tasks, since it causes decoherent effect on investigated quantum system. Nevertheless, the study in recent decades has changed people's view of dissipation, and the environment can be used as a resource
in QIP experimentally \cite{Plenio1999,Beige2000,Horodecki2001,Plenio2002,Yi2003,Diehl2008,Verstraete2009}.

Quantum entanglement, as one of the most striking features in quantum theory, is defined to describe a strongly correlated system constituted
by pairs or groups of particles, and a measurement made on either of the particles
 collapses the state of the system instantaneously. Thus it is an interesting question how can one prepare this kind of strongly `coherent' system using the `decoherent' factors. Currently, there are several representative schemes creating steady bipartite entanglement of high quality by dissipation \cite{Vacanti2009,Kastoryano2011,Busch2011,DallaTorre2013,Rao2013,Carr2013,Leghtas2013,Lin2013,Shankar2013,Bentley2014}. For instance, in the context of cavity QED, the cavity decay was exploited to drive
the system into a maximally entangled stationary state, making the spontaneous emission of atom be the solely detrimental element \cite{Kastoryano2011}.
In neutral atom systems, two groups independently prepared high-fidelity steady-state entanglement between a pair of Rydberg atoms with dissipative Rydberg pumping \cite{Rao2013,Carr2013}. More recently, a highly entangled state with fidelity above $99\%$ was achieved in ion traps, and the fidelity was further enhanced by detection of photons spontaneous emission emitted in to the environment \cite{Bentley2014}. However, it is challenging to detect atomic decay event in experiment.

In this work, we suggest an alternative scheme in the context of cavity QED for dissipatively preparing a maximally entangled state by capitalizing the advantages of above protocols. We combine the spontaneous emission of atom with quantum Zeno dynamics and Rydberg antiblockade effect to drive the system into a purely entangled steady state irrespective of initial state. The concept of quantum Zeno dynamics, formally proposed by Facchi {\it et al} but also adopted by Beige {\it et al} is a generalization of quantum Zeno effect \cite{Facchi2000,Facchi2002,Facchi2008,Beige2000a,Beige2000b,Pachos2002}. The frequent measurements
do not necessarily hinder the evolution of the quantum system
but the system can evolve away from the initial state via
the measurements as long as the projection is multi-dimensional.

For our concerning atom-cavity systems, the quantum Zeno subspaces via a strong continuous coupling is able to effectively suppress the occupation of photon in cavity and has been actively exploited to execute various of QIP based on unitary dynamics \cite{Raimond,Signoles2014,Bretheau2015,Barontini2015}.
The function of Rydberg antiblockade interaction, primarily introduced to break the symmetry by shifting the energy level of triplet states, can also result in an effect of `ground-state blockade' and then speed up the convergence rate of stationary entanglement for certain initial states, as this kind of interaction strength is much larger than the Rabi frequency of microwave field. Generally speaking, regarding two typical noise sources in cavity QED system, we take advantage of atomic decay and avoid the effect of cavity decay at the same time. Thence a fidelity $\sim90\%$ of steady entanglement is available at a low value of single-atom
cooperativity parameter $C\equiv g^2/(\kappa\gamma)=10$. Moreover, by virtue of quantum-jump-based feedback control technology, the same fidelity can be achieved even as $C=5.2$. This feature expands the feasible scope of our scheme from the perspective of experimental realization.

\begin{figure}
\includegraphics[scale=0.13]{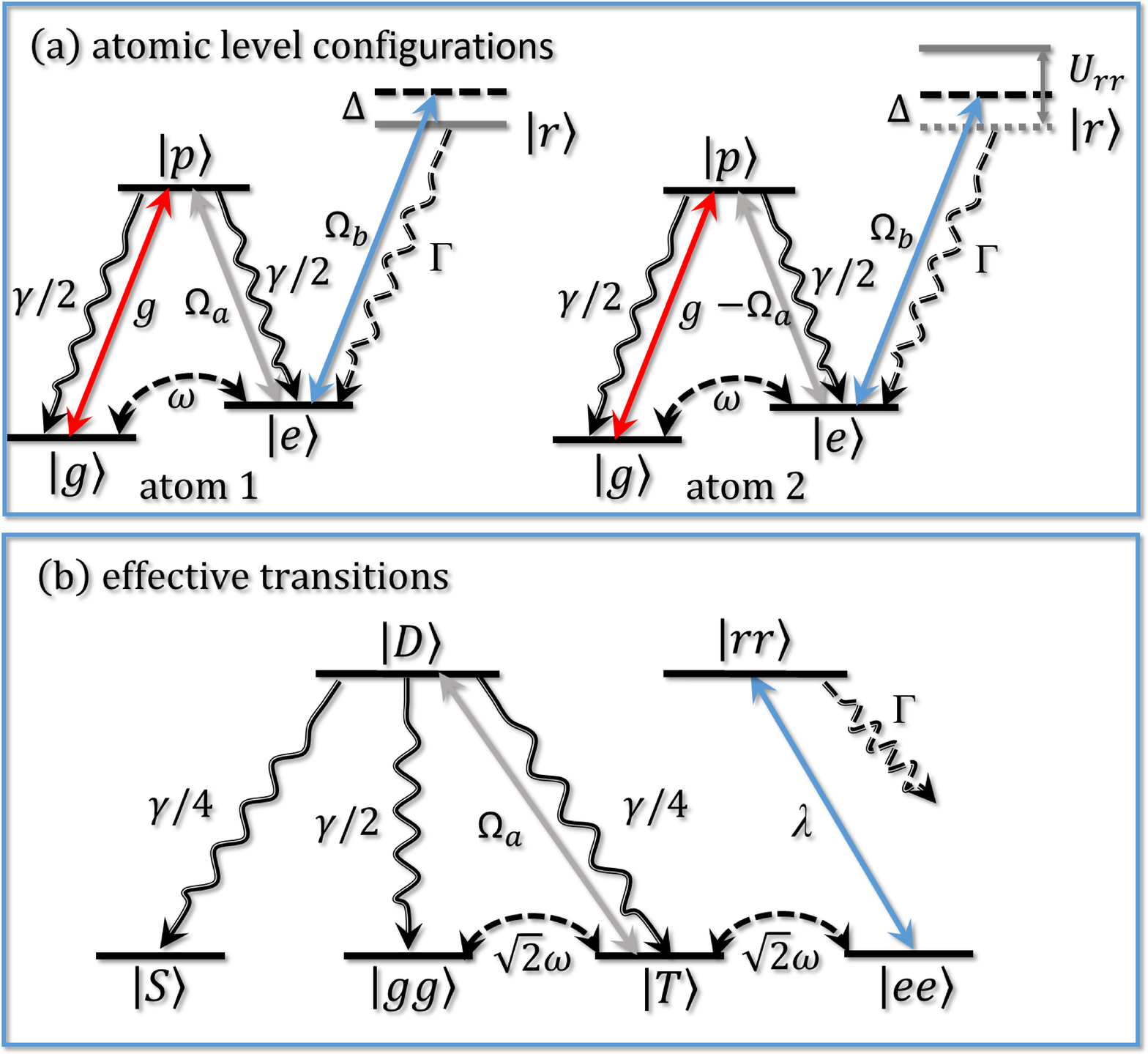}
\caption{\label{fig1} (Color online) (a) Schematic view of atomic-level
configurations. The transition from the ground state $|g\rangle$ to the optical state $|p\rangle$ is resonantly coupled to the quantized cavity field with coupling strength $g$, and the transition between states $|p\rangle$ and $|e\rangle$ is driven by an optical pumping laser acting on the two atoms with Rabi frequency $\pm\Omega_a$. In addition, there is a microwave field $\omega$ causes transitions from the ground state $|g\rangle$ to the ground state $|e\rangle$ which is then pumped to the excited Rydberg state $|r\rangle$ by a classical field $\Omega_b$, detuning $-\Delta$. (b) The effective transitions for two atoms. The whole system works well in a subspace of zero occupation for cavity mode due to the quantum Zeno dynamics. Under the coaction of the spontaneous emission of state $|D\rangle$, the optical pumping laser $\Omega_a$, the microwave driving field $\omega$, and the Rydberg antiblockade interaction $\lambda$, the system will be finally stabilized into the singlet state $|S\rangle$ because of $\Gamma\ll\gamma$. }
\end{figure}
Our system consists of two $N$-type four level Rydberg atoms and an optical cavity, as shown in Fig.~\ref{fig1}(a). A similar configuration of Rydberg atom has been employed to efficiently produce
 multi-particle entangled states using Rydberg blockade interactions by Saffman and M{\o}lmer \cite{Saffman2009}. Here we replace the original Rydberg state $|p\rangle$ with an optical state, since our scheme is completely based on the spontaneous emission of atoms.
 The atoms in ground states $|g\rangle$ and $|e\rangle$ can execute an upward transition to the excited state $|p\rangle$ through resonant interaction with a quantized cavity field of coupling strength $g$, and an optical pumping laser of Rabi frequency $\pm\Omega_a$, respectively.  Apart from this, a microwave field of Rabi frequency $\omega$ is introduced to cause transitions between ground states $|g\rangle$ and $|e\rangle$, and an extra pumping laser field with Rabi frequency $\Omega_b$ drives the atom to the high-lying excited Rydberg state $|r\rangle$ from state $|e\rangle$ detuned by $-\Delta$. For the sake of simplicity, we have assumed the atom decays from the optical state $\ket{p}$ to the ground states $\ket{g}$ and $\ket{e}$ with the same spontaneous emission rate $\gamma/2$, and the life time of the Rydberg state $\ket{r}$ is supposed to be $1/\Gamma$.

The Hamiltonian of the system, in the interaction
picture after performing a rotating with respect to $U=\exp(-i\Delta t\sum_{i=1}^2\ket{r}_{ii}\bra{r})$, can be written as ($\hbar=1$)
\bea\label{e1}
H_I&=&H_z+H_r,\\
H_z &=& \sum_{i=1}^2\bigg[\Omega_a(-1)^{i-1}\ket{p}_{ii}\bra{e} +g\ket{p}_{ii}\bra{g}a+{\rm H.c.}\bigg],\\
H_r&=&\sum_{i=1}^2\bigg[\omega\ket{g}_{ii}\bra{e}+\Omega_b\ket{e}_{ii}\bra{r}+ {\rm H.c.}\bigg]\nonumber\\&&+(U_{rr}-2\Delta)\ket{rr}\bra{rr},\eea
where the Rydberg-mediated interaction $U_{rr}$ originates from the
dipole-dipole potential of the scale $C_3/r^3$ or the long-range van
der Waals interaction proportional to $C_6/r^6$, with $r$ being the
distance between two Rydberg atoms and $C_{3(6)}$ depending on
the quantum numbers of the Rydberg state \cite{Jaksch2000,Urban2009,Mol,Mul,Isenhower2010,Saffman2010,Saffman2016}.

For convenience, we reformulate the Hamiltonian $H_z$ as $H_z=\Omega_a(H_c+KH_q)$, where $K=g/\Omega_a$, $H_c$ stands for the dimensionless  interaction Hamiltonian between atoms and the classical field, and $H_q$ denotes the counterpart between atoms and the quantum cavity field. For a strong coupling limit $K\rightarrow\infty$, the Zeno requirement is satisfied and the above Hamiltonian takes the form \cite{Facchi2002} $H_z=\Omega_a(\sum_nP_nH_cP_n+K\varepsilon_nP_n)$, where $P_n$ is the eigenprojection of the $H_g$ belonging to the
eigenvalue $\varepsilon_n$: $H_q=\sum_n\varepsilon_nP_n$. When we confine our system into the subspace corresponding to $P_0$, the Zeno Hamiltonian reduces to $H_z=\Omega_a\ket{T}\bra{D}\otimes\ket{0}_c\bra{0}+{\rm H.c.}$, with $|T\rangle=(\ket{eg}+\ket{ge})/\sqrt{2}$, and $|D\rangle=(\ket{pg}-\ket{gp})/\sqrt{2}$, $|0_c\rangle$ represents the vacuum state of cavity field. As for the Hamiltonian $H_r$, we can obtain its effective form in the regime of Rydberg antiblockade: $U_{rr}\sim2\Delta\gg\Omega_b$, which is approximated to $H_r=\omega(\ket{g}_{1}\bra{e}+\ket{g}_{2}\bra{e})+\lambda\ket{ee}\bra{rr}+ {\rm H.c.} +(U_{rr}-2\Delta+\lambda)\ket{rr}\bra{rr}$, where
$\lambda=2\Omega_b^2/\Delta$ comes from the second-order perturbation theory, and we also omit the Stark-shift term of level $|e\rangle$. If we further assume that  $(U_{rr}-2\Delta+\lambda)=0$, the Hamiltonian of Eq.~(\ref{e1}) takes the following concise form
\bea\label{e4}
H_I=\Omega_a\ket{T}\bra{D}+\omega\sum_{i=1}^2\ket{g}_{ii}\bra{e}+\lambda\ket{ee}\bra{rr}+ {\rm H.c.}.\eea
Note that the effective model is decoupled to the cavity field due to the restriction of quantum Zeno dynamics.

The Markovian master equation, describing the current system-environment interaction, is modeled in Lindblad form
\be \dot{\rho} = -i[H_I,\rho]+\sum_{j=1}^3 L_j \rho L_j^\dag -\frac{1}{2}(L_j^\dag L_j \rho + \rho L_j^\dag L_j), \label{MastEqn}\ee
and the dominant operators for spontaneous emission read
\be L_{1(2)}=\sqrt{\frac{\gamma}{4}}\ket{S}(\ket{T})\bra{D},\  L_3=\sqrt{\frac{\gamma}{2}}\ket{gg}\bra{D},\ee
where $|S\rangle=(\ket{eg}-\ket{ge})/\sqrt{2}$.
The effective transitions of current reduced system is depicted in  Fig.~\ref{fig1}(b). It can be seen clearly that atoms in the excited state $|D\rangle$ can spontaneously and independently decay into the ground states $\ket{S}$, $\ket{gg}$, and $\ket{T}$, respectively. For the atoms in state $\ket{gg}$ or $\ket{T}$, they will be pumped into other states by the microwave field or the Rydberg antiblockade interaction. Whereas for the atoms staying in state $\ket{S}$, they are invariant on account of $H_I(L_i)\ket{S}=0$. Therefore, the process of pumping and decaying
repeats again and again untill the system is finally stabilized into the state $\ket{S}$.
Although the spontaneous emission of Rydberg state may cause a leakage of quantum information out of related subspace, it contributes little effect on our scheme because $\Gamma\ll\gamma$ generally works for a realistic situation.  In what follows, our simulations are all based on the full Hamiltonian of Eq.~(\ref{e1}) without any specification.
\begin{figure}
\scalebox{0.52}{\includegraphics{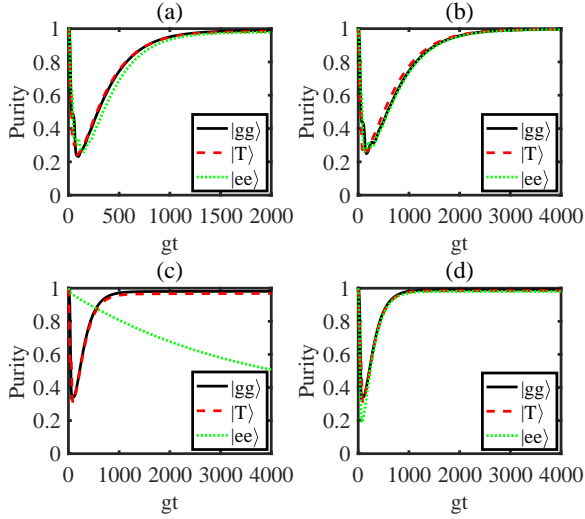} }
\caption{\label{fig2}(Color online) The purity $P(t)=\rm{Tr}[\rho^2(t)]$ is plotted as a function of
time with initial states $|gg\rangle$ (black solid line), $|T\rangle$ (red dashed line) and $|ee\rangle$ (green dotted line), respectively. The spontaneous emission rate of atom is assumed as $\gamma=0.1g$, and other parameters for the calculation
are chosen as (a) $\Omega_a=0.1g$, $\omega=0.05g$, $\Omega_c=0.5g$, $\Delta=10g$; (b) $\Omega_a=0.05g$, $\omega=0.025g$, $\Omega_c=0.5g$, $\Delta=20g$; (c) $\Omega_a=0.05g$, $\omega=0.025g$, $\Omega_c=5g$, $\Delta=100g$; (d) $\Omega_a=0.05g$, $\omega=0.025g$, $\Omega_c=10g$, $\Delta=200g$.}
\end{figure}

In order to fully characterize the dependence of convergence rate on relevant parameters, we use the definition of purity $P(t)=\rm{Tr}[\rho^2(t)]$ and plot its time evolution in Fig.~\ref{fig2} from different initial states. In Fig.~\ref{fig2}(a), we suppose $\Omega_a=0.1g$, which meets the condition of $\Omega_a\ll g$, and other parameter $\omega=0.5\Omega_a$, $\lambda=\omega$ and $\gamma=0.1g$. It shows that all of the initial states tend to be the target state $|S\rangle$ in the same way of evolution and the entangled state is stabilized with $\sim98\%$ purity at $t=2000/g$. In Fig.~\ref{fig2}(b), we reduce half the Rabi frequency  $\Omega_a=0.05g$ so as to satisfy a better Zeno Requirement. Although the convergency time gets longer, the purity of entangled steady state can  reach up to  $99.5\%$. In Fig.~\ref{fig2}(c), we employ $\lambda=20\omega$ and keep other parameters unchanged. Under this condition, the convergence rate for initial states $\ket{gg}$ and $\ket{T}$ is accelerated, accompanied by deceleration for initial state $\ket{ee}$. This result can also be attributed to quantum Zeno effect. Considering the subspace $\{\ket{T},\ket{ee},\ket{rr}\}$, a strong coupling ($\lambda\gg\sqrt{2}\omega$) between $\ket{ee}$ and $\ket{rr}$ has the effect of frequent measurements on the transition $\ket{T}\leftrightarrow\ket{ee}$, thus the mechanism of `ground-state blockade' for occupation of state $\ket{ee}$ is activated for a quantum state initialized in $\ket{gg}$ or $\ket{T}$ \cite{shao}, resulting in a shortcut to the steady state. In contrast, there is only $2\omega^2/\lambda^2$ probability for the initial state $\ket{ee}$ entering into state $\ket{T}$, wherefore
 a much longer time is required for relaxation.
 \begin{figure}
\scalebox{0.50}{\includegraphics{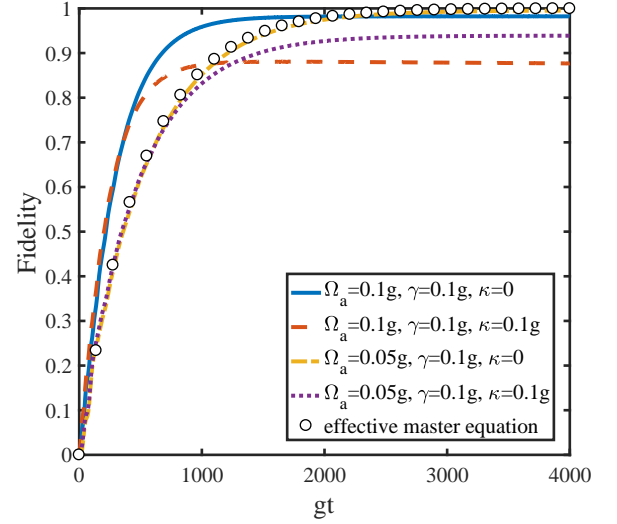} }
\caption{\label{fig3}(Color online) The dependence of the fidelity $F(t)=\bra{S}\rho(t)\ket{S}$ on time is illustrated from the initial state $|gg\rangle$ for a perfect cavity and a leaky cavity, respectively. The other fixed parameters are set as $\omega=0.05g$, $\Omega_c=0.5g$, $\Delta=10g$ for the solid line and dashed line, and $\omega=0.025g$, $\Omega_c=0.5g$, $\Delta=20g$ for the dash-dotted line and dotted line. Compared with the case of $\Omega_a=0.1g$, the fidelity is improved for a weaker driving field $\Omega_a=0.05g$ at the cost of increasing the convergence time. The empty circles simulate the dynamics from the effective master equation corresponding to the dash-dotted line.}
\end{figure}
 Interestingly, as the increase of the ratio $\lambda/\omega$, we find a specific value of $\lambda=g$ that can shorten the convergence time of initial state $\ket{ee}$, too. In Fig.~\ref{fig2}(d), all of the initial states are stabilized into the target state with less time but higher purity than ones in Fig.~2(a). To reveal the physical principle behind this phenomenon, we must take the single excitation from state $\ket{ee}$ into account on the basis of Eq.~(\ref{e4}) as a correction, i.e.
$
H_I=\Omega_a\ket{T}\bra{D}+\omega\sum_{i=1}^2\ket{g}_{ii}\bra{e}+\lambda\ket{ee}\bra{rr}
+\sqrt{2}\Omega_a|ee\rangle\langle B|-g|B\rangle\langle S|a+{\rm H.c.}.$
where $\ket{B}=(\ket{pe}-\ket{ep})/\sqrt{2}$, and the interaction part related to state $\ket{ee}$ could be rewritten as
\bea
H_{ee}=\Omega_a(e^{i\lambda t}\ket{\Phi_+}+e^{-i\lambda t}\ket{\Phi_-})(e^{-igt}\bra{\Psi_+}+e^{igt}\bra{\Psi_-})\eea
with dressed states $\ket{\Phi_\pm}=(\ket{ee}\pm\ket{rr})/\sqrt{2}$, and $\ket{\Psi_\pm}=(\ket{B}\ket{0_c}\mp\ket{S}\ket{1_c})/\sqrt{2}$. Therefore, a fast resonant transition occurs between states $\ket{ee}$ and $\ket{B}$ for $\lambda=g$, which then decays into the stationary entangled state $|S\rangle$ via spontaneous emission.

\begin{figure}
\scalebox{0.40}{\includegraphics{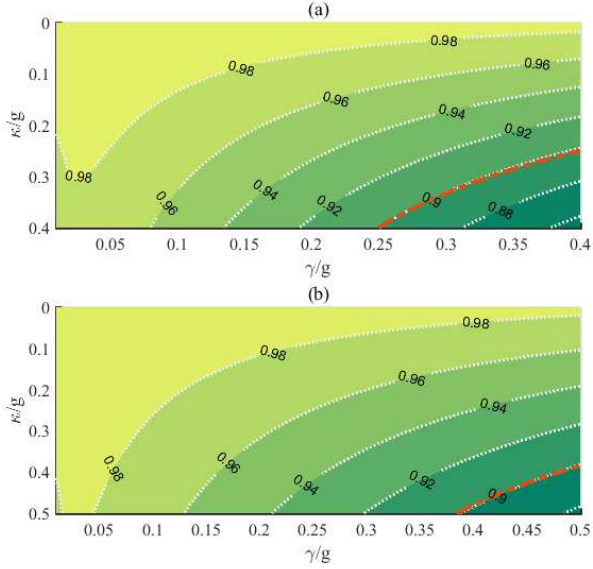} }
\caption{\label{fig4}(Color online) (a) Contour plot of steady-state fidelity  versus atomic spontaneous emission $\gamma/g$ and cavity decay $\kappa/g$ under the given parameters: $\Omega_a=0.01g$, $\omega=0.5\Omega_a$, $\Omega_c=g$, $\Delta=20g$, $\Gamma=0.001g$. The dash-dotted line is simulated as a fidelity around 90\% corresponding to the single-atom cooperativity parameter $C=10$. (b) Contour plot of the enhanced fidelity for feedback-based steady state versus atomic spontaneous emission $\gamma/g$ and cavity decay $\kappa/g$. The feedback parameter is chosen as $\eta=0.5\pi$, and the other relevant parameters are set as $\Omega_a=0.01g$, $\omega=0.5\Omega_a$, $\Omega_c=g$, $\Delta=20g$,  $\Gamma=0.001g$, in order to obtain a 90\% fidelity at a small value of $C=5.2$ (as shown by the dash-dotted line).}
\end{figure}

To show the robustness of our scheme against the leakage of cavity, we introduce the fidelity defined as
$F(t)=\bra{S}\rho(t)\ket{S}$, and simulate its dependence on time from the initial state $\ket{gg}$ with a full master equation including cavity decay $\kappa$ in Fig.~\ref{fig3}. The solid line and the dash-dotted line correspond to the solid lines of Fig.~\ref{fig1}(a) and \ref{fig1}(b), respectively. It can be seen that in the presence of $\kappa=0.1g$, a choice of small $\Omega_a$ favors a high fidelity $93.88\%$ though the convergence time is prolonged, and the effectiveness of quantum Zeno dynamics is again verified.

The above analysis implies that our scheme can actively make use of atomic decay and prevent the malevolent effect of cavity decay simultaneously. Thence it is possible to produce a high-fidelity steady entangled state for a wide range of decoherence parameters. In Fig.~\ref{fig4}(a), the contour of the fidelity for entanglement is plotted after solving the steady-state master equation numerically. The dash-dotted line represent a $\sim90\%$ fidelity corresponding to the single-atom cooperativity parameter $C=10$, which is much smaller than the values required by previous schemes \cite{Kastoryano2011,Busch2011}. In addition, in virtue of quantum-jump-based feedback control, the threshold of $C$ can be further lowered \cite{Carvalho2007,Carvalho2008,Stevenson2011}. In this situation, the feedback master equation is
$
\dot{\rho}=-i[H_I,\rho]+{\cal L_{\rm sp} }\rho+\kappa{\cal D}[{\bf U_{{\rm fb}}}a]\rho,
$
where ${\bf U_{{\rm fb}}}=\exp[-i\eta(\ket{e}_1\bra{g}+\ket{g}_1\bra{e})\otimes {\rm I_2}]$ is the feedback operator acting on the first atom. In Fig.~\ref{fig4}(b), the contours of the enhanced fidelity for feedback-based steady state is simulated and a small value of $C=5.2$ is big enough for the system to keep a $\sim90\%$ fidelity, as illustrated by the dash-dotted line.

\begin{figure}
\scalebox{0.53}{\includegraphics{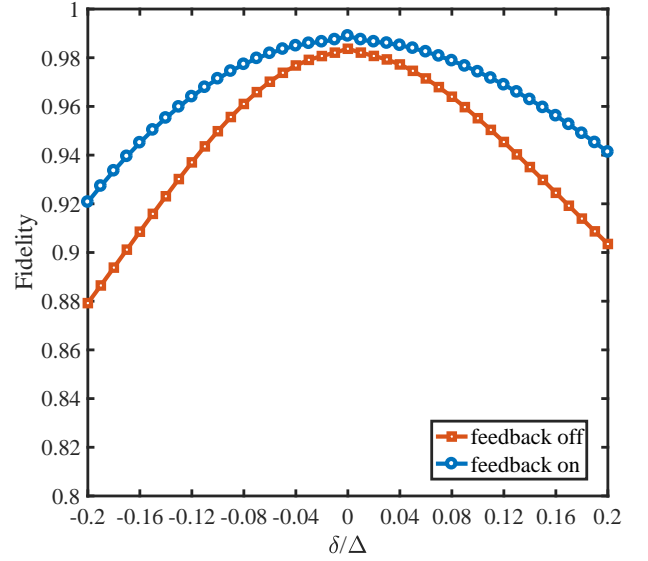} }
\caption{\label{fig5}(Color online) The steady-state fidelity versus the deviation $\delta/\Delta$, where $\kappa=\gamma=0.1g$ is assumed and other parameters are the same as of Fig.~\ref{fig4}.}
\end{figure}

In experiment, the Rydberg antiblockade  condition  $U_{rr}=(2\Delta-\lambda)$ is not readily satisfied, since it is difficult to accurately adjust the distance between two Rydberg atoms. Consequently, a deviation of $U_{rr}$ will deteriorate the quality of final entanglement. In Fig.~\ref{fig5}, we suppose the Rydberg interaction term $U_{rr}=(2\Delta-\lambda+\delta)$ and simulate the steady-state fidelity as a function of the deviation $\delta/\Delta$. Fortunately, the current scheme permits $\delta/\Delta\in(-0.16,0.2]$ so as to preserve the fidelity more than $90\%$. For the current available parameters in the cavity QED with Rydberg-blocked atoms \cite{Brennecke2007,Guerlin2010,Zhang2013,Grankin2014}, the strength coupling the transition between atomic ground level $5S_{1/2}$ and the optical level $5P_{3/2}$ of $^{87}$Rb atom to the quantized cavity mode is $g/2\pi=14.4$ MHz,  the decay rates of the intermediate state $|p\rangle$ is $\gamma_p/2\pi=3$ MHz, the decay rate of the cavity mode is $\kappa/2\pi=0.66$ MHz, and the spontaneous emission rate for the Rydberg state is
$\Gamma/2\pi=1$ KHz. By modulating the Rabi frequencies, detuning and Rydberg interaction strength to be values of Fig.~\ref{fig4}, we obtain the fidelity of steady state $98.6\%$ ($99.0\%$) when the feedback is off (on).

In summary, we have proposed a new mechanism for generation of bipartite entanglement in cavity QED system by dissipation. The convergence time of the stationary entangled state is speeded up via the effect of ground-state blockade and a $\sim90\%$ fidelity can be achieved without resorting to a large value of $C$. Thus our scheme is capable of adapting a wide range of experimental parameters from the perspective of steady state. We hope that this work may open a new venue for the experimental
realization of entanglement in the near future.

This work is supported by the Natural Science Foundation
of China under Grants No. 11647308, No. 11674049,
No. 11534002, and No. 61475033, and by Fundamental
Research Funds for the Central Universities under Grant No.
2412016KJ004.
\bibliographystyle{apsrev4-1}
\bibliography{dissipation_QZD}

\end{document}